\documentclass[aps,pra,reprint,groupedaddress]{revtex4-2}

\usepackage{amsmath}
\usepackage{amssymb}
\usepackage{amsfonts}
\usepackage{amsbsy}
\usepackage{graphicx}
\usepackage{hyperref}
\usepackage[normalem]{ulem}

\hypersetup{
 colorlinks = true, 
 urlcolor = blue, 
 linkcolor = blue, 
 citecolor = blue 
}

\newcommand{\Obs}{\mathbf{O}}
\newcommand{\obs}{\mathcal{O}}
\newcommand{\err}{\epsilon}
\newcommand{\cumErr}{e}
\newcommand{\LLR}{\lambda}
\newcommand{\cumLLR}{l}
\newcommand{\CI}{C}
\newcommand{\BD}{B}
\newcommand{\DR}{G}
\newcommand{\ngPar}{\alpha}
\newcommand{\optPar}{s}
\newcommand{\SDA}{\mathcal{A}}
\newcommand{\gaussPDF}{\phi}
\newcommand{\gaussCDF}{\Phi}
\newcommand{\erfc}{\textrm{erfc}}
\newcommand{\tr}{\textrm{Tr}}
\newcommand{\sgn}{\textrm{sgn}}
\DeclareMathOperator*{\arginf}{\textrm{arginf}}
\newcommand{\bra}[1]{\left<#1\right|}
\newcommand{\ket}[1]{\left|#1\right>}
\newcommand{\ketbra}[2]{\left|#1\right>\!\!\left<#2\right|}
\newcommand{\bs}[1]{\boldsymbol{#1}}

\begin{document}

\title{Generalized figure of merit for qubit readout}

\author{B. D'Anjou}

\email{benjamin.danjou@uni-ulm.de}

\affiliation{Institut f{\"u}r theoretische Physik, Universit{\"a}t Ulm, D-89069 Ulm, Germany}

\date{\today}

\begin{abstract}
Many promising approaches to fault-tolerant quantum computation require repeated quantum nondemolition (QND) readout of binary observables such as quantum bits (qubits). A commonly used figure of merit for readout performance is the error rate for binary assignment in a single repetition. However, it is known that this figure of merit is insufficient. Indeed, real-world readout outcomes are typically analog instead of binary. Binary assignment therefore discards important information on the level of confidence in the analog outcomes. Here, a generalized figure of merit that fully captures the information contained in the analog readout outcomes is suggested. This figure of merit is the Chernoff information associated with the statistics of the analog readout outcomes in one repetition. Unlike the single-repetition error rate, the Chernoff information uniquely determines the asymptotic cumulative error rate for arbitrary readout noise. As a result, non-Gaussian readout noise common in experiments can be described by effective Gaussian noise with the same Chernoff information. Importantly, it is shown that such a universal description persists for the small number of repetitions and non-QND imperfections relevant to real experiments. Finally, the Chernoff information is used to rigorously quantify the amount of information discarded by analog-to-binary conversion. These results provide a unified framework for qubit readout and should facilitate optimization and engineering of near-term quantum devices across all platforms.
\end{abstract}

\maketitle

\section{Introduction}

The ability to readout binary quantum observables such as quantum bits (qubits) is an important {\it desideratum} for quantum information processing~\cite{divincenzo2001}. In particular, it is often highly desirable that the readout have high fidelity and be quantum nondemolition (QND). For instance, many promising fault-tolerant architectures for scalable fault-tolerant quantum computation require that stabilizer parities be repeatedly read out during the computation~\cite{fowler2009,sun2014,kelly2015,cramer2016,ofek2016,rosenblum2018,negnevitsky2018,hu2019,riste2020,andersen2019,andersen2020,bultink2020}. For fault tolerance to be achieved in these architectures, it is crucial that the readout fidelity be above the threshold of the error-correcting code~\cite{knill1998,aharonov2000}. Moreover, it is necessary that the readout be QND so that the code is projected onto the eigenstate corresponding to the observed stabilizer eigenvalues. QND readouts have the important advantage that repeated readouts leave the eigenvalues of the observable unchanged. Therefore, each repetition provides additional information on the observable. As a result, the readout fidelity increases exponentially with the number of repetitions~\cite{deuar1999,deuar2000}. This property has been exploited to improve the readout fidelity of quantum bits (qubits) for a variety of implementations including trapped ion qubits~\cite{schaetz2005,hume2007}, solid-state spin qubits~\cite{meunier2006,jiang2009,neumann2010,robledo2011,waldherr2011,maurer2012,dreau2013,pla2013,lovchinsky2016,boss2017,holzgrafe2019,nakajima2019,yoneda2020,xue2020-2}, and superconducting qubits~\cite{elder2020}. The same temporal correlations in the outcomes of consecutive QND readouts can be used to correct stabilizer readout errors in quantum error-correcting codes~\cite{wang2010,devitt2010,fowler2012-2}.

A seemingly natural figure of merit for the performance of repetitive QND readout is the probability $\err$ of a readout error occurring in a single repetition. Here, each repetition is assigned a binary outcome, with $\err$ being the probability of an incorrect assignment. The readout errors are then corrected by performing a majority vote on the binary outcomes. The cumulative readout error rate $\cumErr_N$ after $N$ repetitions is simply proportional to the probability that an error has occurred in more than half of the repetitions, $\cumErr_N \propto \err^{N/2}$. Therefore, it appears that the cumulative readout error rate is fully determined by the single-repetition error rate $\err$. However, a typical real-world readout does not only have two outcomes. Rather, the readout outcomes are commonly analog (see Fig.~\ref{fig:fig1}) and need not even be scalar. For instance, the single-repetition readout outcome could be a continuous electrical voltage or current~\cite{elzerman2004,barthel2009,mallet2009,morello2010,studenikin2012,jeffrey2014,saira2014,broome2017,walter2017,nakajima2017,pakkiam2018,vukusic2018,harveycollard2018,opremcak2018,west2019,urdampilleta2019,zheng2019,keith2019,keith2019-2,opremcak2021,ebel2020,connors2020,martinez2020,rosenthal2021,jang2020}, a nonbinary photon count at a photodetector~\cite{myerson2008,gehr2010,robledo2011,harty2014,shields2015,danjou2016,hopper2020,todaro2021,edmunds2020}, or a collection of such outcomes. If each individual repetition is assigned a binary outcome, information on the level of confidence in each analog outcome is discarded. Such analog-to-binary conversion is known as ``hard decoding''. It was shown that taking into account the additional information contained in the distribution of analog readout outcomes, or ``soft decoding'', can significantly reduce $\cumErr_N$ compared to hard decoding~\cite{danjou2014-2,hann2018,dinani2019,liu2020,xue2020-2}. It follows that two repetitive QND readouts characterized by the same value of $\err$ can yield different values of $\cumErr_N$. This suggests that $\err$ is not a universal descriptor of readout performance~\footnote{The importance of soft decoding was also recognized in the context of continuous-variable quantum error correction~\cite{fukui2017,vuillot2019,noh2020-3,noh2020,rozpedek2020,fukui2020}, continuous-variable quantum communication~\cite{bari2012}, and quantum parameter estimation~\cite{danjou2014-2,ryan2015,xu2019-7}.}. Moreover, it was shown that the existence of a soft-decoding advantage is highly dependent on the details of the often highly non-Gaussian distributions of analog readout outcomes. Heuristic arguments have been put forward to predict when an advantage exists in common cases~\cite{danjou2014-2,xue2020-2}, but a unified and economical description that fully captures the performance of repetitive QND readout for all outcome distributions is highly desirable.

The present work suggests a figure of merit that fully captures the cumulative error rate $\cumErr_N$ of the repetitive QND readout of binary observables with an arbitrary distribution of analog readout outcomes. That figure of merit is the asymptotic rate of decrease of $\ln \cumErr_N$ with the number of repetitions $N$. In the classical theory of hypothesis testing, this quantity is known as the Chernoff information for the discrimination of two probability distributions~\cite{chernoff1952,hoeffding1965}. Like the single-repetition readout error rate $\err$, the Chernoff information can be obtained solely from the statistics of readout outcomes in a single repetition. In fact, it is closely related to $\err$ when the readout outcomes are binary. Unlike $\err$, however, the Chernoff information does not discard information associated with the level of confidence in each analog readout outcome. Therefore, the Chernoff information enables a universal description of repetitive QND readout, in the sense that all outcome distributions with the same Chernoff information have the same asymptotic cumulative readout fidelity. Therefore, theoretical analysis of repetitive QND readout is reduced to the calculation of the Chernoff information. Importantly, this universality persists in the nonasymptotic regime and in the presence of non-QND imperfections. This leads to simple and universal expressions for the cumulative error rate of a QND readout that remain accurate for small $N$. Finally, the Chernoff information is used to predict the soft-decoding advantage in cases of practical importance without having to resort to time-consuming simulations~\cite{danjou2014-2,hann2018,dinani2019,nakajima2019,liu2020,xue2020-2}. The present work paves the way for a generalized understanding of the real-world readout of quantum observables and should facilitate the engineering of high-fidelity QND readout in near-term quantum devices on all platforms.

\section{Repetitive quantum nondemolition readout \label{sec:qndReadoutRepetitive}}

\subsection{Quantum nondemolition readout \label{sec:qndReadout}}

A binary quantum observable $A$ has only two distinct eigenvalues $a = +1$ and $a = -1$. The observable $A$ could be, e.g., the $Z$ Pauli observable of a qubit or a parity-check stabilizer in an error-correcting code. If the system is prepared in an eigenstate of $A$ with eigenvalue $a$, an ideal QND readout of $A$ yields the value $a$ with certainty. Moreover, the post-readout state is also an eigenstate with eigenvalue $a$. However, a real-world QND readout is subject to noise that introduces uncertainty in the value of $a$. In general, it is therefore not possible to determine $a$ with certainty after a single readout. Fortunately, the QND property guarantees that every subsequent readout yields the same eigenvalue as in the first readout. Thus, repeated readouts ``average out'' the noise and enable readout of the observable $A$ to arbitrary accuracy. A more detailed overview of the theory of quantum nondemolition readout is given in Appendix~\ref{app:qndReadout}.

\subsection{Single repetition \label{sec:singleRepetition}}

\begin{figure}
	\centering
	\includegraphics[width=\columnwidth]{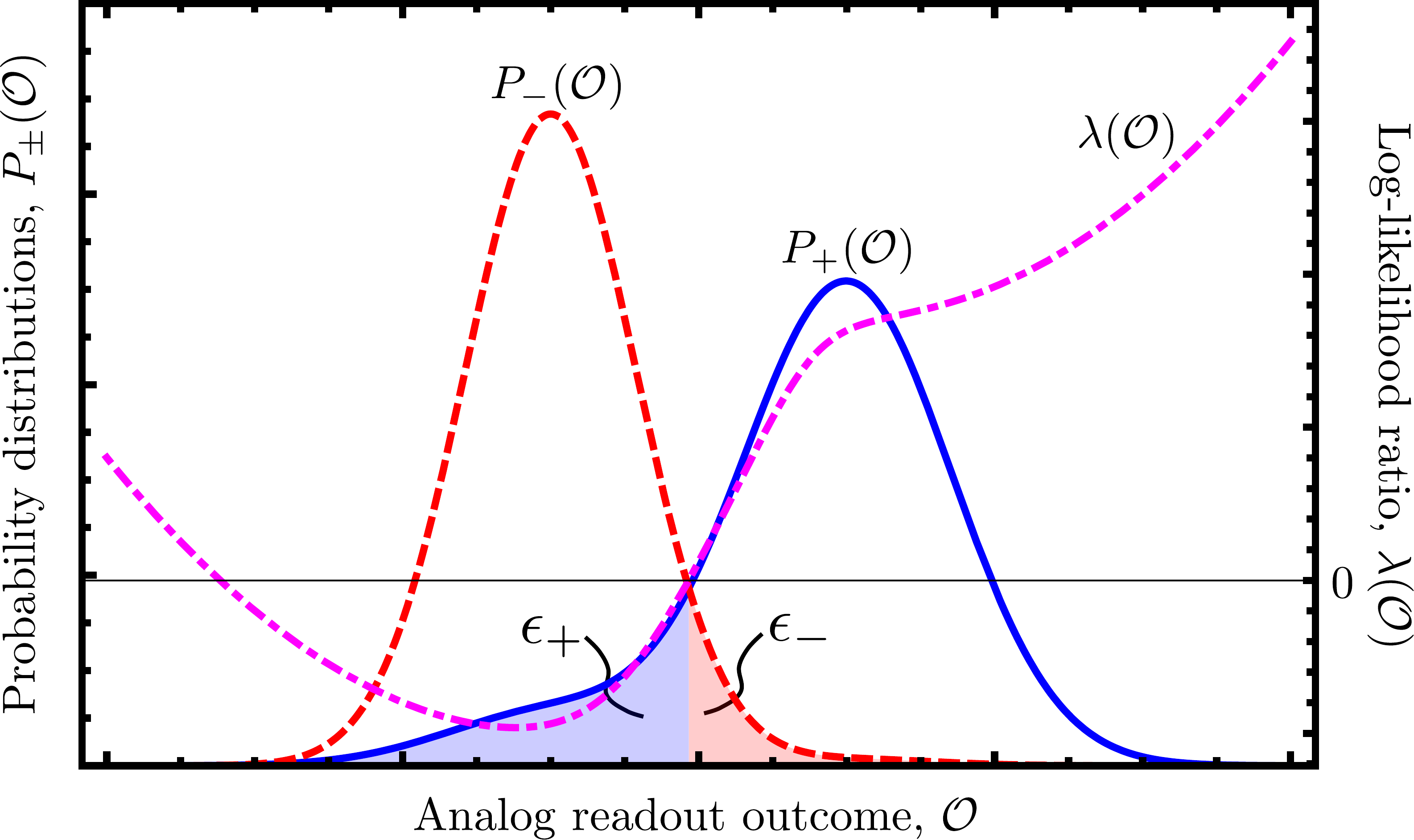}
	\caption{Schematic illustration of single-repetition probability distributions $P_\pm(\obs)$ of analog readout outcomes conditioned on the observable eigenvalue $a=\pm 1$. The corresponding log-likelihood ratio $\LLR(\obs)$ is also shown. The single-repetition error rates $\err_\pm$ conditioned on the eigenvalue $a = \pm 1$ are represented by the shaded areas. \label{fig:fig1}}
\end{figure}

A single repetition of a general QND readout yields an outcome $\obs$ that depends on the eigenvalue $a$. More precisely, the statistics of the readout outcomes are described by the probability distribution $P_\pm(\obs)$ for observing $\obs$ if $a = \pm 1$. Here, the distributions $P_\pm(\obs)$ can take any form. For instance, the outcome $\obs$ could be a discrete random variable, a continuous random variable, or a multidimensional set of random variables. Possible distributions $P_\pm(\obs)$ are illustrated schematically in Fig.~\ref{fig:fig1}. Several other experimentally relevant examples are discussed in Secs.~\ref{sec:generalized} and \ref{sec:softDecoding}. In the following, it is assumed that these distributions are known \emph{a priori}, either empirically or from theoretical modeling. The precise meaning of the distributions $P_\pm(\obs)$ within quantum measurement theory is reviewed in Appendix~\ref{app:qndReadout}.

The most commonly used figure of merit for readout performance in a single repetition is the error rate $\err$, defined as the average probability of assigning the incorrect eigenvalue to the observed outcome. The value of $\err$ depends on the rule chosen to assign an eigenvalue $a$ to each outcome $\obs$. In the following, it is assumed that the two eigenvalues are equally likely {\it a priori}. This leads to a definition of $\err$ that is agnostic about the value of $a$. Moreover, this case is common and desirable because it maximizes the information extracted by readout. Under this assumption, the assignment rule that minimizes $\err$ is obtained by calculating the log-likelihood ratio~\cite{kay1998}
\begin{align}
	\LLR(\obs) = \ln \frac{P_+(\obs)}{P_-(\obs)}. \label{eq:LLR}
\end{align}
 When $\LLR(\obs)$ is larger (smaller) than $0$, the eigenvalue $a = +1$ ($a = -1$) is assigned. If $\LLR(\obs) = 0$, the eigenvalue is assigned at random. The log-likelihood ratio, Eq.~\eqref{eq:LLR}, is central to hypothesis testing. It should be interpreted as the observer's level of confidence in the assignment given the observed outcome $\obs$. The log-likelihood ratio is depicted in Fig.~\ref{fig:fig1} alongside the distributions $P_\pm(\obs)$. The average single-repetition error rate is $\err = \left(\err_+ + \err_-\right)/2$, where
\begin{align}
		\err_+ = P_+(\LLR < 0), \;\;\;\err_- = P_-(\LLR > 0) \label{eq:conditionedErrors}
\end{align}
are the error rates conditioned on preparation of $a = +1$ and $a = -1$, respectively. Here, $P_\pm(\LLR)$ are the probability distributions for $\LLR$ conditioned on $a = \pm 1$. The conditioned error rates $\err_\pm$ are represented by the shaded areas~\footnote{Or hypervolumes in the case of multidimensional $\obs$.} in Fig.~\ref{fig:fig1}. Because $\err_+$ and $\err_-$ are, respectively, integrals of $P_+(\obs)$ and $P_-(\obs)$ only, the error rate $\err$ cannot contain information on the relative value of $P_+(\obs)$ and $P_-(\obs)$. Therefore, important information contained in the functional form of log-likelihood ratio $\LLR(\obs)$ is discarded.

\subsection{Multiple repetitions \label{sec:multipleRepetitions}}

That the single-repetition error rate discards information is most readily seen by considering the repeated QND readout of the binary observable $A$. Repeated readout yields a string of outcomes $\Obs_N = \left\{\obs_0,\obs_1,\dots,\obs_{N-1}\right\}$. Due to the QND nature of the readout, all outcomes are independently sampled from the same distribution $P_\pm(\obs)$ when an eigenstate with eigenvalue $a$ is prepared. Accordingly, the joint distribution of the readout outcomes conditioned on the eigenvalue $a = \pm 1$ is $P_\pm(\Obs_N) = \prod_{k=0}^{N-1} P_\pm(\obs_k)$. A quantum-mechanical derivation is given in Appendix~\ref{app:qndReadout}. The cumulative log-likelihood ratio for the entire string of outcomes $\Obs_N$ is thus
\begin{align}
	\cumLLR_N = \ln \frac{P_+(\Obs_N)}{P_-(\Obs_N)} = \sum_{k=0}^{N-1} \LLR(\obs_k). \label{eq:cumLLR}
\end{align}
As before, the eigenvalue $a = +1$ ($a = -1$) is assigned when $\cumLLR_N > 0$ ($\cumLLR_N < 0$). Equation~\eqref{eq:cumLLR} shows that in the general case, each outcome must be weighed by $\LLR(\obs_k)$ in order to perform optimal assignment. Therefore, discarding information contained in $\LLR(\obs)$ in each repetition is necessarily suboptimal. The average cumulative error rate is now $\cumErr_N = (\cumErr_{+,N}+\cumErr_{-,N})/2$, where
\begin{align}
	\cumErr_{+,N} = P_+(\cumLLR_N < 0), \;\;\;\cumErr_{-,N} = P_-(\cumLLR_N > 0) \label{eq:cumConditionedErrors}
\end{align}
are the cumulative error rates conditioned on preparation of $a = +1$ and $a = -1$, respectively. Here, $P_\pm(\cumLLR_N)$ are the probability distributions for $\cumLLR_N$ conditioned on $a = \pm 1$. Because the noise is sampled independently in each repetition, $\cumErr_N$ is expected to decrease exponentially as $N$ grows, $\cumErr_N \propto \exp\left(-\CI N\right)$ for some constant $\CI$. The constant $\CI$ is the Chernoff information, which will be argued to be a more appropriate figure of merit for repetitive QND readout than the single-repetition error rate $\err$.

\section{A generalized figure of merit \label{sec:generalized}}

\subsection{The Chernoff information \label{sec:chernoffInformation}}

The asymptotic behavior of the cumulative error rate $\cumErr_N$ is given by the theory of large deviations~\cite{cover2005} developed by Cram{\' e}r~\cite{cramer1938,*cramer1994,*cramer2018} and Sanov~\cite{sanov1957,*sanov1961,*sanov1961-2}, and applied to hypothesis testing by Chernoff and Hoeffding~\cite{chernoff1952,hoeffding1965}. The theory is summarized in Appendix~\ref{app:largeDeviationTheory}. The result is that
\begin{align}
\begin{array}{lcl}
\ln \cumErr_N \sim - \CI N & \textrm{as} & N\rightarrow \infty,
\end{array} \label{eq:exponentialDependence}
\end{align}
where
\begin{align}
	\CI = -\inf_{\optPar\in\left[0,1\right]} \ln \left[ \int d\obs\,P_+(\obs)^\optPar P_-(\obs)^{1-\optPar} \right]. \label{eq:chernoffInformation}
\end{align}
Here, ``$\sim$'' denotes asymptotic equality. The quantity $\CI$ is known as the Chernoff information~\footnote{In the literature, the Chernoff information is also known as the Chernoff bound or the Chernoff distance. A quantum version of the Chernoff information has also been developed~\cite{audenaert2007,audenaert2008,calsamiglia2008,nussbaum2009} which optimizes the asymptotic cumulative error rate over all possible quantum measurements. However, the present work is concerned with readout performance for a given imperfect local measurement. The classical Chernoff information is sufficient to that end.}. It is a symmetric distance measure between the distributions $P_+(\obs)$ and $P_-(\obs)$ and can be interpreted as a rate of information gain per repetition. Like the single-repetition readout error rate $\err$, the Chernoff information depends only on the statistics of readout outcomes in a single repetition. Unlike $\err$, however, it depends on the relative value of $P_+(\obs)$ and $P_-(\obs)$. Consequently, the Chernoff information encodes information contained in the level of confidence $\LLR(\obs)$ in each readout outcome. As a result, readout outcome distributions $P_\pm(\obs)$ with the same single-repetition error rates $\err_\pm$ do not necessarily have the same Chernoff information.

\subsection{Universality \label{sec:universalityQND}}

\begin{figure*}
	\centering
	\includegraphics[width=\textwidth]{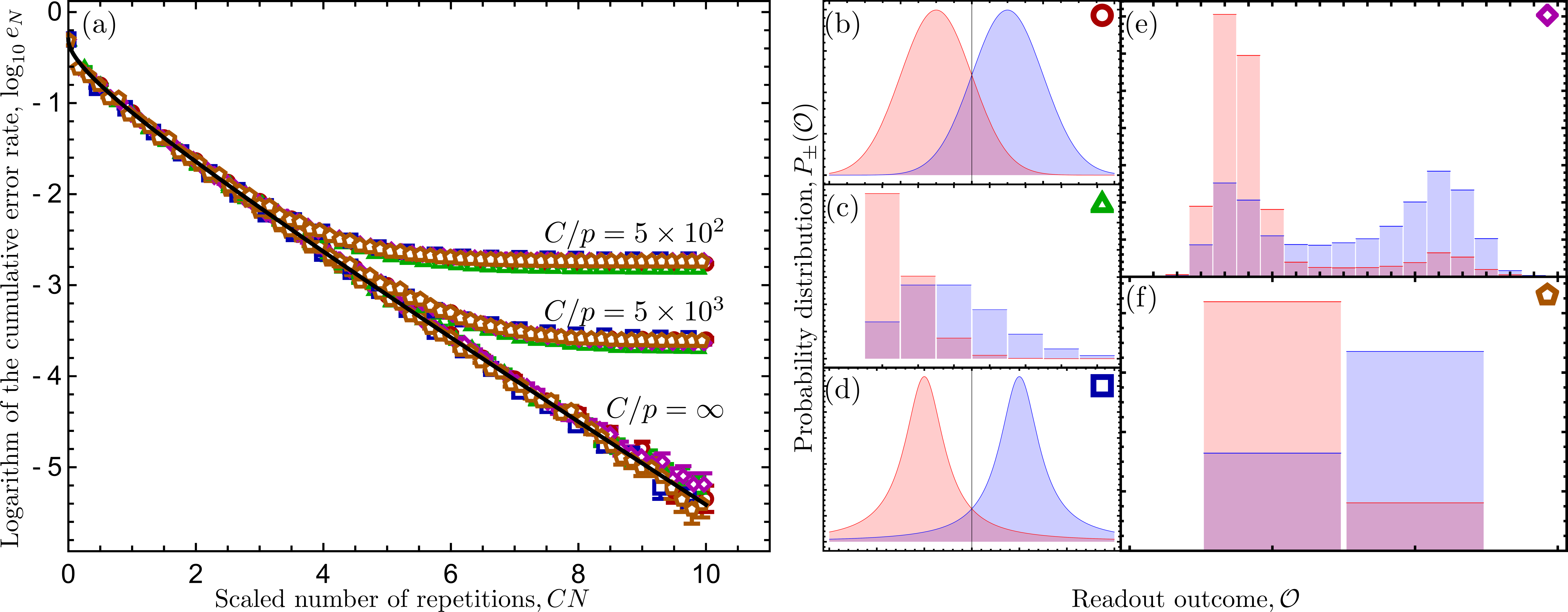}
	\caption{(a) Monte Carlo simulations of the cumulative error rate $\cumErr_N$ for (b) Gaussian noise ($\CI = 0.5$, $\ngPar = 1$, $\optPar^* = 0.5$), (c) Poissonian noise ($\CI = 0.2533$, $\ngPar = 0.9999, \optPar^* = 0.5569$), (d) Cauchy noise ($\CI = 0.4422$, $\ngPar = 1.1079$, $\optPar^* = 0.5$), and (e),(f) the analog ($\CI = 0.1634$, $\ngPar = 1.0522$, $\optPar^* = 0.5203$) and binary ($\CI = 0.1577$, $\ngPar = 1.0536$, $\optPar^* = 0.5199$) noise observed in Ref.~\cite{xue2020-2}. In (b)-(f), the blue distribution corresponds to $a = +1$ and the red distribution corresponds to $a = -1$. The cumulative error rate is a universal function of $\CI N$ and $\CI/p$ for all simulated noise models. The ideal QND case corresponds to $\CI/p = \infty$. The solid black line is obtained from Eq.~\eqref{eq:gaussianAnsatz}. The details of the simulations are discussed in Appendix~\ref{app:simulations}. \label{fig:fig2}}
\end{figure*}

The power of using the Chernoff information as a figure of merit for readout of binary observables is that all readout outcome distributions $P_\pm(\obs)$ with the same Chernoff information, no matter their shape, give the same asymptotic behavior for $\ln\err_N$ as $N\rightarrow \infty$. Here, it is argued that such universal behavior persists even in the nonasymptotic regime $N \gtrsim 1$. As discussed in Sec.~\ref{sec:gaussianDistributions}, the Chernoff information for Gaussian noise with signal-to-noise ratio $r$ is simply given by $\CI = r/2$. This suggests an interpretation of the Chernoff information as an effective Gaussian signal-to-noise ratio. More precisely, it suggests that as far as the cumulative error rate $\cumErr_N$ is concerned, non-Gaussian noise may be replaced by an effective Gaussian noise with signal-to-noise ratio $2\CI$. In the case of Gaussian noise, however, an exact expression for $\cumErr_N$ can be obtained for all $N$, namely, $\cumErr_N = \cumErr_{\pm,N} = \erfc\left(\sqrt{r N/2}\right)/2$~\cite{gambetta2007}. This naturally leads to the ansatz that the same relationship holds for arbitrary noise by setting $r = 2\CI$:
\begin{align}
	\cumErr_N = \cumErr_{\pm,N} = \frac{1}{2}\erfc\left(\sqrt{\CI N}\right). \label{eq:gaussianAnsatz}
\end{align}
It can be shown from simple counter-examples at $N=1$ that Eq.~\eqref{eq:gaussianAnsatz} is not exact for finite $N$. Nevertheless, its approximate validity was assessed by performing numerical Monte Carlo simulations for a variety of readout outcome distributions $P_\pm(\obs)$ (see Appendix~\ref{app:simulations}). The results are shown in Fig.~\ref{fig:fig2}. It is found that Eq.~\eqref{eq:gaussianAnsatz} captures $\ln \cumErr_N$ extremely well for all $N \ge 1$ and for all the considered noise models. These include Gaussian noise ubiquitous in readouts relying on electronic and homodyne detection~\cite{barthel2009,morello2010,jeffrey2014,saira2014,nakajima2017,broome2017,walter2017,harveycollard2018,pakkiam2018,west2019,urdampilleta2019,zheng2019,keith2019,connors2020,opremcak2021,rosenthal2021,jang2020}, Poissonian noise ubiquitous in readouts relying on fluorescence detection~\cite{myerson2008,gehr2010,robledo2011,harty2014,shields2015,danjou2016,hopper2020,todaro2021,edmunds2020}, Cauchy noise with fat polynomial tails, and the heavily bimodal non-Gaussian readout noise observed empirically in Ref.~\cite{xue2020-2}. The latter two cases show that Eq.~\eqref{eq:gaussianAnsatz} approximately holds for noise distributions that are heavily non-Gaussian and need not even have a finite variance~\footnote{Some of the qubit readout schemes cited here are not QND. Nevertheless, such qubits may still be used as ancillas to perform QND readout of other observables. In such cases, the non-Gaussian features of the noise translate directly to the repetitive QND readout discussed in this manuscript.}. The approximate validity of Eq.~\eqref{eq:gaussianAnsatz} can be intuitively understood with the following argument. For a fixed value of $\CI N$, the number of readouts $N$ increases as $\CI \rightarrow 0$. Therefore, the noise becomes effectively Gaussian on all time scales and the limit of a continuous readout subject to Gaussian noise is recovered~\cite{gambetta2007,tsang2012}. This expresses a generalized central-limit theorem for the probability of rare events in Eq.~\eqref{eq:cumConditionedErrors}. What the simulations in Fig.~\ref{fig:fig2} shows is that Eq.~\eqref{eq:gaussianAnsatz} remains an excellent approximation for common non-Gaussian sources of noise with finite $N$ and finite $\CI \lesssim 1$ . This is precisely the situation where repetitive QND readout is most useful.

In the regime where Eq.~\eqref{eq:gaussianAnsatz} is less accurate, $\CI \gtrsim 1$, a more general approximate universal form of the cumulative error rates is obtained with the help of a saddle-point approximation~\cite{vantrees2001,butler2015}. While such an approximation becomes more accurate as $N$ increases, it typically remains very accurate for finite $N$~\cite{butler2015}. As discussed in Appendix~\ref{app:largeDeviationTheory}, the saddle-point approximation for the average error rate is
\begin{align}
\begin{split}
	\cumErr_N \approx \frac{1}{2}&\erfc\left(\sqrt{\CI N}\right) + \frac{\left(\ngPar^{-1/2}-1\right)}{\sqrt{4\pi\CI N}}\exp\left(-\CI N\right), \label{eq:saddlePointLeading}
\end{split}
\end{align}
and the saddle-point approximations for the conditioned error rates are
\begin{align}
	\cumErr_{\pm,N} \approx \cumErr_N \pm \frac{\left(2\optPar^* - 1\right)}{\sqrt{4\pi \ngPar \CI N}}\exp\left(-\CI N\right). \label{eq:saddlePointLeadingConditioned}
\end{align}
Here, $\ngPar$ is a parameter that is easily computed from the distributions $P_\pm(\obs)$ as described in Appendix~\ref{app:largeDeviationTheory}. Moreover, $\optPar^*$ is the position of the optimum in Eq.~\eqref{eq:chernoffInformation}. These parameters quantify the deviation from the Gaussian behavior of Eq.~\eqref{eq:gaussianAnsatz}. Indeed, it is shown in Appendix~\ref{app:largeDeviationTheory} that $\ngPar \rightarrow 1$ and $\optPar^* \rightarrow 1/2$ as $\CI \rightarrow 0$. In that limit, the saddle-point approximation reduces to Eq.~\eqref{eq:gaussianAnsatz} for all $\CI N$. Note that universality persists even in the extreme case where the saddle-point approximation breaks down, $\ngPar \CI N \ll 1$. In that case, the cumulative error rate approaches the Chernoff upper bound, $\cumErr_N \approx \exp\left(-\CI N\right)/2$~\cite{cover2005}. Finally, it must be noted that deviations from Eqs.~\eqref{eq:saddlePointLeading} and \eqref{eq:saddlePointLeadingConditioned} may occur at finite $N$ for discrete readout noise~\footnote{More precisely, Eqs.~\eqref{eq:saddlePointLeading} and \eqref{eq:saddlePointLeadingConditioned} must be modified if the cumulative log-likelihood ratio $\cumLLR_N$ takes a discrete set of values for some finite $N$. This can occur even for some continuous distributions $P_\pm(\obs)$. For instance, the single-repetition log-likelihood ratio $\LLR(\obs)$ is approximately binary-valued for the near-continuous distributions in Fig.~\ref{fig:fig2}(e). Finally, note that exact analytical expressions can be obtained for $\cumErr_N$ and $\cumErr_{\pm,N}$ in the special cases of Poissonian noise and binary noise.}. It is possible to modify the above expressions to account for these so-called ``continuity corrections'' if necessary~\cite{butler2015}.

The above discussion shows that Eq.~\eqref{eq:gaussianAnsatz} can be used to accurately estimate the cumulative error rate $\cumErr_N$ for arbitrary analog readout noise and finite $N$, obviating the need for time-consuming simulations~\cite{danjou2014-2,hann2018,dinani2019,nakajima2019,liu2020,xue2020-2} that are specialized to the noise model. Even in the regime where Eq.~\eqref{eq:gaussianAnsatz} becomes less accurate, universal behavior is retained at the cost of introducing only two additional parameters $\ngPar$ and $\optPar^*$. This approximate nonasymptotic universal behavior should greatly facilitate readout engineering by reducing the analysis of the great variety of noise models discussed in the literature to the calculation of the Chernoff information and its auxiliary quantities $\ngPar$ and $\optPar^*$.

\subsection{Non-QND imperfections \label{sec:universalityNonQND}}

In practice, the error rate of a QND readout is limited by non-QND imperfections that generate transitions between the eigenvalues of $A$. To achieve a low cumulative error rate $\cumErr_N$, non-QND processes must necessarily act on a time scale longer than (1) the duration $\Delta t$ of a single readout and longer than (2) the time scale $\Delta t/\CI$ for achieving low error rate. In this ``single-shot readout'' regime, transitions between the eigenvalues of $A$ are effectively classical~\cite{gagen1993,korotkov2001}. More precisely, the observed quantum jumps are well described by classical transition probabilities for all Markovian non-QND processes (see Appendix~\ref{app:qndReadout} for a more detailed discussion). A very common case is relaxation from $a=+1$ to $a=-1$ with small probability $p \ll \min(\CI,1)$ in each repetition~\cite{myerson2008,barthel2009,jeffrey2014,saira2014,walter2017,pakkiam2018,west2019,urdampilleta2019,zheng2019,nakajima2019,yoneda2020,xue2020-2,connors2020,opremcak2021,rosenthal2021}. It is important to verify that universality persists in this more realistic scenario. 

For Gaussian noise, it is known that $\cumErr_N$ is a function of $r N$ and $r/p$ only~\cite{gambetta2007}. Therefore, $\cumErr_N$ should be a universal function of $\CI N$ and $\CI/p$ regardless of the details of the noise in the regime where Eq.~\eqref{eq:gaussianAnsatz} holds, $\ngPar \approx 1$ and $\optPar^* \approx 1/2$. It was verified that this is indeed the case by performing Monte Carlo simulations using the procedure described in Appendix~\ref{app:simulations}. The results are shown in Fig.~\ref{fig:fig2}. All noise models with the same value of $\CI/p$ collapse on the same curve when plotted as a function of $\CI N$. Therefore, the cumulative error rate $\cumErr_N$ may simply be tabulated for various values $\CI N$ and $\CI/p$ by assuming Gaussian noise. The cumulative error rate for arbitrary non-Gaussian noise can then be directly read off from the Gaussian results. Although the simulation is not shown here, it was also verified that the above conclusions hold when both relaxation and excitation occur with probability $p$.

In the regime where Eq.~\eqref{eq:gaussianAnsatz} is less accurate, $\ngPar \neq 1$ and $\optPar^* \neq 1/2$, it was observed numerically that the above conclusions hold provided that the values of $\ngPar$ and $\optPar^*$ are also specified. That is, the logarithm of the cumulative readout error rate appears to have the functional form
\begin{align}
\ln \cumErr_N = f\left(\CI N, \CI/p, \ngPar, \optPar^* \right). \label{eq:nonQNDUniversality}
\end{align}
Note that the average error rate may now depend on $\optPar^*$ because non-QND imperfections may affect different states asymmetrically. Just as in the perfectly QND case, additional ``continuity corrections'' may be required in the case of discrete distributions and finite $N$~\cite{butler2015}.

\section{Hard and soft decoding \label{sec:softDecoding}}

\subsection{Soft-decoding advantage \label{sec:softDecodingAdvantage}}

The Chernoff information can also be used to quantify the information lost by converting analog readout outcomes to binary values. To do this, the Chernoff information $\CI$ for analog outcomes is compared to the Chernoff information $\CI_b$ for the corresponding binarized outcomes. This is reflected in the soft-decoding advantage
\begin{align}
	\SDA = \frac{\CI}{\CI_b}. \label{eq:advantage}
\end{align}
If $\SDA = 1$, no information is discarded by binarizing readout outcomes. If $\SDA > 1$, however, a significant amount of information has been lost. Indeed, inspection of Eq.~\eqref{eq:exponentialDependence} shows that binarizing readout outcomes changes the order of magnitude of $\cumErr_N$ by a factor $\SDA$,
\begin{align}
	\cumErr_N \propto \left(\cumErr_{N,b}\right)^\SDA \;\;\; \textrm{as} \;\;\; N \rightarrow \infty. \label{eq:orderMagnitude}
\end{align}
Here, $\cumErr_{N,b}$ is the cumulative error rate for binarized outcomes. Equivalently, the number of readouts required to achieve a desired value of $\cumErr_N$ is $\SDA$ times larger with hard decoding than with soft decoding. Accounting for such lost information could prove critical in pushing readout errors below the threshold of quantum error-correcting codes. Due to the persistence of universality at small $N$ discussed in Sec.~\ref{sec:universalityQND}, the asymptotic soft-decoding advantage is expected to also persist in the nonasymptotic limit. Indeed, it was verified that Eq.~\eqref{eq:advantage} accurately predicts the soft-decoding advantage observed for small $N$ in Ref.~\cite{xue2020-2}.

A general analytical expression for $\CI_b$ is given in Appendix~\ref{app:softDecodingAdvantage}. In the important limit $\err_\pm \rightarrow 0$, it takes the form
\begin{align}
	\CI_b \sim \left[\frac{1}{\ln(\err_+^{-1})}+\frac{1}{\ln(\err_-^{-1})}\right]^{-1}. \label{eq:asymptoticBinaryChernoffInformation}
\end{align}
Similar to the average single-repetition error rate, $\err = (\err_+ + \err_-)/2$, $\CI_b$ is a monotonic function of both $\err_+$ and $\err_-$. This makes $\CI_b$ an appropriate substitute for $\err$ to quantify the performance of a single repetition.

To illustrate the usefulness of the Chernoff information in characterizing readout, the soft-decoding advantage, Eq.~\eqref{eq:advantage}, is now calculated for two examples of experimental interest.

\subsection{Example 1: Gaussian distributed readout outcomes \label{sec:gaussianDistributions}}

It is first assumed that the readout of the eigenvalues $a = \pm 1$ is subject to additive Gaussian noise, such that the distributions of analog readout outcomes are
\begin{align}
 P_\pm(\obs) = \sqrt{\frac{r}{2\pi}} \exp\left[-\frac{r\left(\obs \mp 1\right)^2}{2}\right]. \label{eq:distsGaussian}
\end{align}
Here, $r$ is the (power) signal-to-noise ratio. Gaussian noise is ubiquitous in real experiments. For instance, electronic noise in the readout of semiconductor spin qubits~\cite{barthel2009,morello2010,nakajima2017,broome2017,harveycollard2018,pakkiam2018,west2019,urdampilleta2019,zheng2019,keith2019,connors2020,jang2020} as well as quantum noise in the readout of superconducting qubits~\cite{jeffrey2014,saira2014,walter2017,opremcak2021,rosenthal2021} are well modeled by additive Gaussian noise. An application of Eq.~\eqref{eq:chernoffInformation} gives $\CI = r/2$ for all $r$. The single-repetition error rates corresponding to these distributions are $\err_\pm = \erfc{\left(\sqrt{r/2}\right)}/2$. Using this expression, it is possible to show that $\CI_b \approx r/\pi$ for $r \ll 1$ and $\CI_b \approx r/4$ for $r \gg 1$ (see Appendix~\ref{app:softDecodingAdvantage}). Thus, the soft-decoding advantage varies smoothly from
\begin{align}
\begin{array}{lcl}
\SDA = \frac{\pi}{2} \textrm{ for } r \ll 1& \textrm{ to } &\SDA = 2 \textrm{ for } r \gg 1
\end{array}.
\end{align}
Therefore, hard decoding of Gaussian distributions leads to loss of information for all signal-to-noise ratios. In particular, the number of repetitions required to reach a desired error rate is always at least $\pi/2 \approx 1.57$ times larger if the analog outcomes are binarized. The result for $r\gg1$ is consistent with the analysis of Ref.~\cite{danjou2014-2} and with known results from the classical theory of soft-decision decoding~\cite{chase1972,einarsson1976}.

\subsection{Example 2: Gaussian distributed readout outcomes with conversion errors \label{sec:conversionErrors}}

\begin{figure}
	\centering
	\includegraphics[width=\columnwidth]{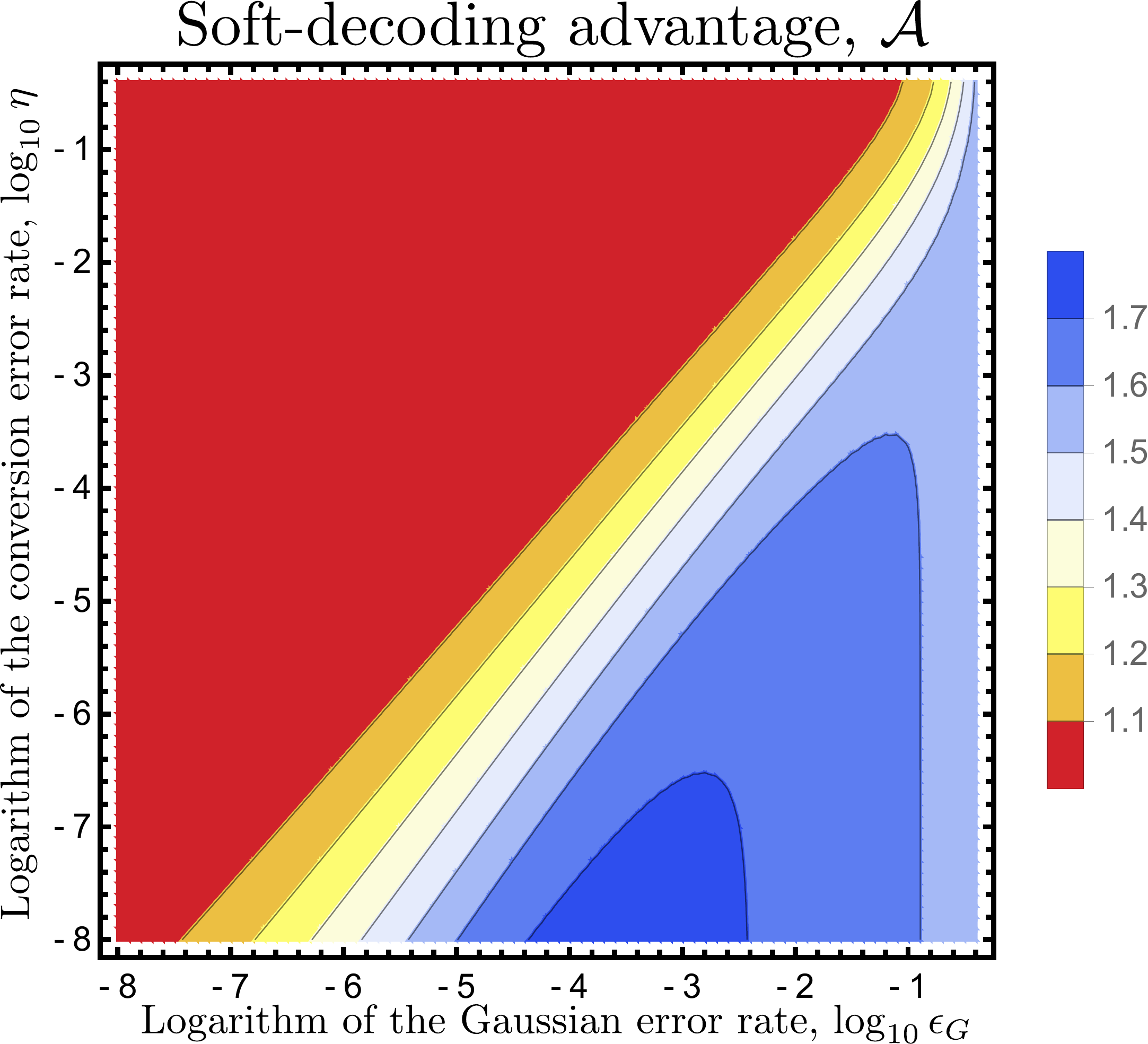}
	\caption{Soft-decoding advantage $\SDA$ for a QND readout with both Gaussian noise and conversion errors. Here, $\err_G = \erfc\left(\sqrt{r/2}\right)/2$ is the rate of pure Gaussian errors and $\eta$ is the rate of conversion errors. Note that the Chernoff information increases as $\err_G$ and $\eta$ decrease. The soft-decoding advantage was calculated using the expressions given in Appendix~\ref{app:softDecodingAdvantage}. \label{fig:fig3}}
\end{figure}

In the presence of Gaussian readout noise, the eigenvalues $a = \pm 1$ are ideally each converted to Gaussian distributions with means $\pm 1$. In practice, however, imperfections in the readout scheme may lead to conversion errors. As a result, the distributions $P_\pm(\obs)$ often resemble mixtures of Gaussian distributions~\cite{morello2010,barthel2009,jeffrey2014,saira2014,walter2017,pakkiam2018,west2019,urdampilleta2019,zheng2019,xue2020-2}. Such imperfections can be modeled with the distributions
\begin{align}
\begin{split}
P_\pm(\obs) = &(1-\eta)\, \sqrt{\frac{r}{2\pi}} \exp\left[-\frac{r\left(\obs \mp 1\right)^2}{2}\right]\\
  &+ \eta\, \sqrt{\frac{r}{2\pi}} \exp\left[-\frac{r\left(\obs \pm 1\right)^2}{2}\right].
\end{split} \label{eq:distsConversion}
\end{align}
Here, $\eta$ is the rate of conversion errors. Expressions for $\CI$ and $\CI_b$ for these distributions are given in Appendix~\ref{app:softDecodingAdvantage}. The resulting soft-decoding advantage $\SDA$ is shown in Fig.~\ref{fig:fig3} as a function of $\eta$ and of the rate of errors due to Gaussian noise, $\err_G \equiv \erfc{\left(\sqrt{r/2}\right)}/2$.
There is a clear transition from a region of parameter space where $\SDA = 1$ when conversion errors dominate to a region where $\SDA > 1$ when Gaussian errors dominate. This agrees with the heuristic conclusions of Refs.~\cite{danjou2014-2,xue2020-2}. Note, however, that while previous work had to resort to time-consuming simulations to quantify soft-decoding advantage of non-Gaussian distributions~\cite{danjou2014-2,hann2018,dinani2019,nakajima2019,liu2020,xue2020-2}, the present approach enables an accurate prediction of $\SDA$ by computing a single integral. This makes it much easier to explore the parameter space to engineer and optimize readout.

\section{Conclusion}

In conclusion, a generalized figure of merit for the repetitive QND readout of a binary quantum observable $A$ was suggested. This figure of merit is the Chernoff information associated with the analog distributions of readout outcomes for each eigenvalue of $A$ [see Eq.~\eqref{eq:chernoffInformation}]. When the readout outcomes are binary, the Chernoff information is closely related to the commonly used single-repetition error rate. Contrary to the single-repetition error rate, however, the Chernoff information is a universal figure of merit: all noise models with the same Chernoff information yield the same asymptotic functional form for the cumulative error rate. It follows that arbitrary non-Gaussian readout noise can be modeled by effective Gaussian noise without loss of generality. Crucially, it was shown that universal behavior persists for the small number of repetitions and non-QND imperfections relevant to real-world experiments. Finally, the Chernoff information was used to quantify the amount of information discarded by binarizing readout outcomes in each repetition, and simple results were derived analytically for experimentally relevant readout models. The results presented here provide a unified description of repetitive QND readout and should greatly facilitate the standardization, optimization, and engineering of quantum readout across all experimental platforms.

There are several possible avenues for future research. Firstly, it would be interesting to generalize the results presented here for noise that is nonstationary or correlated between repetitions, two features that are likely to appear in real experiments. Extensions to the discrimination of nonbinary observables~\cite{leang1997,li2016} could also be of interest in some architectures. Moreover, a rigorous justification of the universal behavior described by Eq.~\eqref{eq:nonQNDUniversality} is highly desirable. Such a justification could potentially be obtained by using extensions of large deviation theory for Markov processes~\cite{donsker1976,butler2015} or through the relationship between probability theory and the renormalization group~\cite{jonalasinio2001}. In the latter approach, the readout outcomes $\obs_k$ are thought of as classical degrees of freedom on an $N$-site lattice and non-QND transition probabilities are interpreted as weak interaction parameters between the $\obs_k$. The functional form of the error rate could then be obtained by analyzing the renormalization group flow around the fixed point at $N=\infty$. While the present work focuses on repetitive QND readout, the results are expected to be relevant for all quantum information processing tasks where large streams of analog readout outcomes must be processed. In particular, it would be of great interest to investigate whether the Chernoff information can help to universally parametrize the logical failure rate~\cite{iyer2018} as well as quantify the soft-decoding advantage for quantum error-correcting codes subject to arbitrary readout noise.

\begin{acknowledgments}
This work was supported by the \href{https://www.bmbf.de/}{BMBF} project DiaPol, the \href{https://www.dfg.de/}{DFG} via a Reinhart Koselleck award, and the \href{https://erc.europa.eu/}{ERC} Synergy Grant HyperQ. The author thanks W.A.~Coish, M.B.~Plenio, X.~Xue, and M.C.~Korzeczek for useful comments on the manuscript, as well as X.~Xue, T.F.~Watson, and L.M.K.~Vandersypen for sharing data.
\end{acknowledgments}

\appendix

\section{QUANTUM NONDEMOLITION READOUT \label{app:qndReadout}}

\subsection{Review of quantum measurement theory}

Let $\varrho_0$ be the density operator of a quantum system prior to readout. The system is assumed to be initially disentangled from the environment. The density operator $\varrho_1$ conditioned on a readout outcome $\obs$ is then~\cite{wiseman2010}:
\begin{align}
	\varrho_1(\obs) = \frac{\sum_\mu M_\mu(\obs)\varrho_0 M^\dagger_\mu(\obs)}{P(\obs)} \equiv \frac{\mathcal{M}(\obs)\circ \varrho_0}{P(\obs)}. \label{eq:quantumOperation}
\end{align}
Here, the $M_\mu(\obs)$ are the Kraus operators for outcome $\obs$ and $P(\obs)$ is the probability of outcome $\obs$ occurring. The Kraus operators define a linear quantum operation $\mathcal{M}(\obs)$ on density operators (indicated with the symbol ``$\circ$''). The quantum operation depends on the details of the readout apparatus and on the initial state of the environment. The probability $P(\obs)$ is expressed in terms of the positive operator-valued measure (POVM) elements $E(\obs) = \sum_\mu M_\mu^\dagger(\obs) M_\mu(\obs)$:
\begin{align}
	P(\obs) = \tr\left[E(\obs)\varrho_0\right].
\end{align}
The unconditioned density operator $\rho$ is the average of $\varrho$ over all readout outcomes:
\begin{align}
	\rho = \int d\obs\, P(\obs) \varrho(\obs). \label{eq:unconditionedDO}
\end{align}

\subsection{Quantum nondemolition readout of a binary observable}

The readout of an observable $A$ is said to be quantum nondemolition (QND) if the interaction of the system with the readout apparatus commutes with $A$. This implies that the Kraus operators $M_\mu(\obs)$ and $M_\mu^\dagger(\obs)$ commute with each other and with $A$. Therefore,
\begin{align}
	M_\mu(\obs) = \sum_x m_{\mu,x}(\obs)\ketbra{x}{x}, \label{eq:krausQND}
\end{align}
where the $\left\{\ket{x}\right\}$ are common eigenstates of the Kraus operators and of $A$. The POVM elements for the readout of Eq.~\eqref{eq:krausQND} are
\begin{align}
	E(\obs) = \sum_x P_x(\obs)\ketbra{x}{x},
\end{align}
where the distributions of outcomes for each eigenstate are
\begin{align}
	P_x(\obs) = \sum_\mu |m_{\mu,x}(\obs)|^2.
\end{align}
For the readout to be ``of the observable $A$'' only, the outcome distributions must be the same for eigenstates $\ket{x}$ that share the same eigenvalue~\cite{wiseman2010}, $P_x(\obs) = P_{a(x)}(\obs)$. The binary observable discussed in the main text has only two eigenvalues $a = \pm 1$. In this case, the POVM elements take the form
\begin{align}
	E(\obs) = P_+(\obs) \Pi_+ + P_-(\obs) \Pi_-. \label{eq:singlePOVM}
\end{align}
Here, $\Pi_\pm$ is the projector on the eigenspace of $A$ with eigenvalue $a = \pm 1$. The distributions $P_\pm(\obs)$ depend on the details of the readout. They must be normalized but can otherwise take any form. The resulting distribution of outcomes is
\begin{align}
	P(\obs) = P_+(\obs)P(+) + P_-(\obs)P(-),
\end{align}
where $P(\pm) = \tr\left(\Pi_\pm \varrho_0\right)$ are the initial occupations of the eigenspaces of $A$.

Using Eqs.~\eqref{eq:quantumOperation}, \eqref{eq:unconditionedDO}, and \eqref{eq:krausQND}, it is simple to verify that the QND readout does not change the unconditioned occupations of the eigenspaces of $A$:
\begin{align}
	\tr\left[\Pi_\pm \rho_1\right] = \tr\left[\Pi_\pm \rho_0\right]. \label{eq:unconditionedOccupation}
\end{align}
Moreover, let $\ket{x_+}$ and $\ket{x_-}$ be two eigenstates that correspond to distinct eigenvalues of $A$. The magnitude of the coherence between these two eigenstates after readout is
\begin{align}
	\left|\bra{x_+}\rho_1\ket{x_-}\right| = e^{-\DR}\left|\bra{x_+}\rho_0\ket{x_-}\right|, \label{eq:unconditionedCoherence}
\end{align}
where the rate of decoherence per readout is
\begin{align}
	\DR = -\ln \left|\int d\obs\, \sum_\mu m_{\mu,x_+}(\obs)m_{\mu,x_-}^*(\obs)\right|.
\end{align}
Applying the triangle inequality for the integral followed by the Cauchy-Schwarz inequality for the sum yields
\begin{align}
	\DR \geq \BD,
\end{align}
where
\begin{align}
	\BD = -\ln \left[\int d\obs\,\sqrt{P_+(\obs)P_-(\obs)}\right]
\end{align}
is the Bhattacharyya distance between the distributions $P_+(\obs)$ and $P_-(\obs)$. Note that the Chernoff information gives both an upper and a lower bound to the Bhattacharyya distance, $\CI/2 \leq \BD \leq \CI$~\footnote{The upper bound follows from the definition of the Chernoff information as $\CI = \sup_{\optPar\in[0,1]}c(\optPar)$, with $c(\optPar) = -\ln \left[\int d\obs P_+(\obs)^\optPar P_-(\obs)^{1-\optPar}\right]$. Indeed, $C \geq c(1/2) = \BD$. The lower bound follows from noting that the function $c(\optPar)$ is concave and that therefore its supremum $\CI$ at some $\optPar^*\in[0,1]$ must lie below the line $y(\optPar)=2\BD \optPar$ [$y(\optPar)=2\BD(1-\optPar)$] when $\optPar^* > 1/2$ [$\optPar^* < 1/2$]. Thus, $\CI \leq 2\BD$ regardless of the value of $\optPar^*$.}. Therefore, discriminating the eigenvalues $a = \pm 1$ of $A$ guarantees decoherence between the eigenspaces of $A$ as expected~\cite{korotkov1999,korotkov2001,korotkov2001-2}:
\begin{align}
	\DR \geq \CI/2. \label{eq:decoherenceChernoff}
\end{align}

\subsection{Repetitive quantum nondemolition readout of a binary observable}

Repeating the readout $N$ times yields a string of outcomes $\Obs_N = \left\{\obs_0,\obs_1,\dots,\obs_{N-1}\right\}$. If the environment and readout apparatus are reset to their initial state after each readout (the Markov assumption), the density operator $\varrho_N$ conditioned on the outcome of $N$ repetitions is obtained by repeated applications of Eq.~\eqref{eq:quantumOperation}:
\begin{align}
\begin{split}
	\varrho_N = \frac{\mathcal{M}_{N-1}(\obs_{N-1})\circ \dots \circ \mathcal{M}_0(\obs_0)\circ \varrho_0}{P(\Obs_N)}.
\end{split}
\end{align}
The probability distribution of the string $\Obs_N$ is $P(\Obs_N) = \tr\left[E(\Obs_N)\varrho_0\right]$, with the cumulative POVM element
\begin{widetext}
\begin{align}
	E(\Obs_N) = \sum_{\mu_{N-1},\dots,\mu_0} M^\dagger_{\mu_0}(\obs_0)\dots M^\dagger_{\mu_{N-1}}(\obs_{N-1})M_{\mu_{N-1}}(\obs_{N-1}) \dots M_{\mu_0}(\obs_0). \label{eq:povmMultiple}
\end{align}
\end{widetext}
For a QND readout, all Kraus operators $M_\mu(\obs)$ and $M_\mu^\dagger(\obs)$ commute with each other and with the POVM elements $E(\obs)$. Using this property and Eq.~\eqref{eq:povmMultiple} yields $P(\Obs_N) = \tr\left[E(\obs_{N-1})\dots E(\obs_0)\varrho_0\right]$. Using Eq.~\eqref{eq:singlePOVM} then leads to
\begin{align}
\begin{split}
	P(\Obs_N) = P_+&(\obs_{N-1})\dots P_+(\obs_0) P(+)\\
			 &+ P_-(\obs_{N-1})\dots P_-(\obs_0) P(-). \label{eq:classicalStatistics}
\end{split}
\end{align}
The coherences between eigenspaces of $A$ do not enter Eq.~\eqref{eq:classicalStatistics}. Thus, the statistics of QND readout outcomes are identical to the statistics for the repeated readout of a binary classical observable with state-dependent noise $P_\pm(\obs)$. The likelihood ratio of Eq.~\eqref{eq:cumLLR} is therefore the appropriate readout statistic. Moreover, note that the dynamics of the unconditioned density operator is simply given by repeated application of Eqs.~\eqref{eq:unconditionedOccupation} and \eqref{eq:unconditionedCoherence}. The occupations remain unchanged for all $N$, $\tr\left[\Pi_\pm \rho_N\right] = \tr\left[\Pi_\pm \rho_0\right]$, and the coherences decay exponentially at rate $\DR$, $\left|\bra{x_+}\rho_N\ket{x_-}\right| = e^{-\DR N}\left|\bra{x_+}\rho_0\ket{x_-}\right|$.

\subsection{Non-QND imperfections}

In real experiments, there are usually other physical processes that do not conserve the binary observable $A$. In the following, these processes are assumed to be Markovian, i.e., their correlation time is much smaller than the duration of a single readout. The analysis of non-Markovian processes goes beyond the scope of this work.

To simplify the discussion, the case where the observable is the Pauli $Z$ observable of a single qubit, $A = Z$, is considered first. For a perfectly QND readout, the common eigenstates $\left\{\ket{x}\right\}$ of the Kraus operators and of $A$ are simply the eigenstates $\ket{\pm}$ of $Z$. The presence of a non-QND Markov process causes transitions between $\ket{+}$ and $\ket{-}$. This could be due to, e.g., a $T_1$ relaxation process, pure dephasing in a basis other than $\ket{\pm}$, coherent gate errors, or unwanted readout backaction. The repetitive readout can only improve readout if the non-QND process acts on a time scale $t_D$ that is long compared to 1) the duration $\Delta t$ of a single readout and to 2) the time scale $\Delta t/\CI$ required to achieve a low error rate:
\begin{align}
	t_D \gg \frac{\Delta t}{\min(\CI,1)}. \label{eq:singleShotReadoutLimit}
\end{align}
Equation~\eqref{eq:singleShotReadoutLimit} defines the single-shot readout regime. The condition $t_D \gg \Delta t$ ensures that the distributions of outcomes $P_\pm(\obs)$ for each individual readout remain unchanged by the transitions. Moreover, the condition $t_D \gg \Delta t/\CI$ guarantees that the occupations of the eigenstates change slowly compared to the coherence between them [see Eq.~\eqref{eq:decoherenceChernoff}]. Therefore, the coherences quickly reach a steady state that depends only on the instantaneous occupations of the eigenstates $\ket{\pm}$. As a result, the coherences can be adiabatically eliminated from the equations of motion to yield effective classical rate equations for the occupations. More precisely, the observed quantum jumps take the form of a random telegraph signal~\cite{gagen1993,korotkov2001} with transition rates $\lesssim 1/t_D$. This is the physics of the quantum Zeno effect. In this limit, a classical hidden Markov model such as the one discussed in Appendix~\ref{app:simulations} is sufficient to accurately describe the statistics of the readout outcomes.

For a more general binary observable $A$ with degenerate eigenspaces, the general argument is slightly complicated by the fact that there can be coherent dynamics within the degenerate eigenspaces during readout. In practice, however, readout is often designed so that the observed quantum trajectories are effectively classical. In quantum stabilizer codes, notably, a series of parity-check stabilizers $\left\{A_0,A_1,A_2,\dots\right\}$ encode logical qubits in degenerate subspaces with fixed syndrome eigenvalues $\left\{a_0,a_1,a_2,\dots\right\}$. The stabilizers are chosen so that (logical) error processes acting directly within these subspaces occur extremely rarely during one round of stabilizer readout. The dominant error processes are the ones that change the syndrome~\cite{nielsen2010}. In the single-shot readout limit, Eq.~\eqref{eq:singleShotReadoutLimit}, all stabilizers are read out rapidly compared to the time scale of these errors. Using the same argument as for the single-qubit case, the coherences between eigenspaces with different syndromes can be adiabatically eliminated from the dynamics. The observed quantum trajectories then take the form of classical trajectories in the space of syndromes. From the point of view of a single stabilizer, say $A_0$, the observed trajectories are also classical, though the statistics of the associated quantum jumps are not necessarily described by a simple random telegraph signal.

\section{LARGE DEVIATION THEORY \label{app:largeDeviationTheory}}

\subsection{The necessity of large deviation theory}

The cumulative log-likelihood ratio $\cumLLR_N$, Eq.~\eqref{eq:cumLLR}, is the sum of the independent and identically distributed (i.i.d.) variables $\LLR(\obs_k)$. According to the central-limit theorem, the distributions $P_\pm(\cumLLR_N)$ therefore asymptotically converge to Gaussians. Thus, one might hope to evaluate the cumulative error rates, Eq.~\eqref{eq:cumConditionedErrors}, using the cumulative distribution function of a Gaussian distribution. As was first noted by Cram{\'e}r, however, this approach produces wildly inaccurate results~\cite{cramer1938,*cramer1994,*cramer2018}. Indeed, according to the Berry-Esseen theorem~\cite{berry1941,esseen1942}, the cumulative distribution function of $P_\pm(\cumLLR_N)$ converges only polynomially to a Gaussian as $N$ increases. Meanwhile, the error rates $\cumErr_{\pm,N}$ decrease exponentially with $N$. Therefore, the relative accuracy in $\cumErr_{\pm,N}$ explodes as $N\rightarrow \infty$ and the central-limit theorem fails. This problem is solved with the theory of large deviations summarized below.

\subsection{Asymptotic large deviations}

Let $\bar{x} = (1/N) \sum_{k=0}^{N-1} x_k$ be the sample mean of $N$ i.i.d. variables $x_k$. In its simplest form, the main result of large deviation theory is that the complementary cumulative distribution function of $\bar{x}$ asymptotically satisfies~\cite{cover2005}
\begin{align}
	\ln P(\bar{x} > x) \sim - N I(x), \label{eq:sanovTheorem}
\end{align}
where the so-called rate function $I(x)$ is independent of $N$. Here, ``$\sim$'' denotes asymptotic equality as $N\rightarrow \infty$. Large deviation theory provides an explicit expression for the function $I(x)$:
\begin{align}
	I(x) = -\inf_\optPar \left[K(\optPar)-x\,\optPar\right]. \label{eq:cramerRate}
\end{align}
Here, $K(\optPar)$ is the cumulant-generating function of the distribution $P(x_k)$:
\begin{align}
	K(\optPar) = \ln \left[\int ds \,P(x)\exp(x \optPar)\right]. \label{eq:CGF}
\end{align}

Setting $\bar{x} = \cumLLR_N/N$ and $x = 0$ yields asymptotic expressions for the conditioned cumulative error rates, Eq.~\eqref{eq:cumConditionedErrors}:
\begin{align}
	\ln \cumErr_{\pm,N} \sim - \CI_\pm N.
\end{align}
Here,
\begin{align}
	\CI_\pm = -\inf_{\optPar\in\left[0,1\right]} K_\pm(\mp \optPar), \label{eq:sanovTheoremLLR}
\end{align}
and
\begin{align}
	K_\pm(\mp\optPar) = \ln \left[ \int d\LLR P_\pm(\LLR) \exp(\mp \LLR \optPar) \right]. \label{eq:CGFLLR}
\end{align}
Note that in Eq.~\eqref{eq:sanovTheoremLLR}, the infimum occurs in the interval $\optPar\in\left[0,1\right]$. This is because the cumulant-generating functions $K_\pm(\mp\optPar)$ are convex and vanish at $\optPar=0$ and $\optPar=1$. Therefore, the infimum must occur between $\optPar=0$ and $\optPar=1$. Finally, the integration variable is changed back to the readout outcome $\obs$ by setting $\lambda = \ln \left[P_+(\obs)/P_-(\obs)\right]$ in Eq.~\eqref{eq:CGFLLR}. This yields
\begin{align}
\begin{split}
	&K_+(-\optPar) = \ln\left[\int d\obs\, P_+(\obs)^{1-\optPar} P_-(\obs)^\optPar\right],\\
	&K_-(+\optPar) = \ln\left[\int d\obs\, P_+(\obs)^\optPar P_-(\obs)^{1-\optPar}\right].
\end{split} \label{eq:CGFLLRExplicit}
\end{align}
These functions have the same infimum. Therefore, $\CI_+ = \CI_- = \CI$, with $\CI$ given by Eq.~\eqref{eq:chernoffInformation}.

\subsection{Nonasymptotic corrections}

The nonasymptotic corrections to Eq.~\eqref{eq:sanovTheorem} are most naturally obtained using a saddle-point expansion of the cumulant-generating function $K(\optPar)$. It yields an approximate expression for the complementary cumulative distribution function of a continuous-valued sample mean $\bar{x} = (1/N)\sum_{k=0}^{N-1} x_k$~\cite{butler2015}:
\begin{align}
	P(\bar{x} > x) \approx \gaussCDF(-w_N) + \gaussPDF(w_N)\left(\frac{1}{u_N}-\frac{1}{w_N}\right).
\end{align}
Here, $\gaussCDF(z) = \erfc\left(-z/\sqrt{2}\right)/2$ and $\gaussPDF(z) = \exp\left(-z^2/2\right)/\sqrt{2\pi}$ are the standard normal cumulative distribution function and normal probability density function, respectively, and
\begin{align}
\begin{split}
	&w_N = \sgn(\optPar^*)\sqrt{2N\left[x \optPar^* - K(\optPar^*)\right]},\\
	&u_N = \optPar^* \sqrt{N K''(\optPar^*)}.
\end{split}
\end{align}
In this expression, $\optPar^*$ is the same optimum as in Eq.~\eqref{eq:cramerRate}:
\begin{align}
	\optPar^* = \arginf_\optPar \left[K(\optPar) - x \optPar \right] \Leftrightarrow K'(\optPar^*) = x. \label{eq:saddlePointEquation}
\end{align}
The condition in Eq.~\eqref{eq:saddlePointEquation} is known as the saddle-point equation. Applying the above expressions to the cumulant-generating functions $K_\pm(\mp s)$ of the log-likelihood ratio by setting $\bar{x} = \cumLLR_N/N$ and $x=0$ gives the conditioned and average cumulative error rates:
\begin{align}
\begin{split}
	&\cumErr_{\pm,N} \approx \cumErr_N \pm \frac{\left(2\optPar^* - 1\right)}{\sqrt{4\pi \ngPar \CI N}}\exp\left(-\CI N\right), \\
	&\cumErr_N \approx \frac{1}{2}\erfc\left(\sqrt{\CI N}\right) + \frac{\left(\ngPar^{-1/2}-1\right)}{\sqrt{4\pi\CI N}}\exp\left(-\CI N\right).
\end{split} \label{eq:saddlePointErrorRates}
\end{align}
Here,
\begin{align}
	\ngPar = \frac{2{\optPar^*}^2(1-\optPar^*)^2 K_-''(\optPar^*)}{\CI}.
\end{align}
The values of $\optPar^*$ and $\CI = -K_-(\optPar^*)$ are obtained either by minimizing $K_-(\optPar)$ directly in Eq.~\eqref{eq:CGFLLRExplicit} or by solving the saddle-point equation $K_-'(\optPar^*) = 0$. In terms of the readout noise distributions $P_\pm(\obs)$, the saddle-point equation takes the form
\begin{align}
	\int d\obs\, P_\textrm{eff}(\obs) \LLR(\obs) = 0,
\end{align}
where $\LLR(\obs) = \ln \left[P_+(\obs)/P_-(\obs)\right]$ and where the following effective distribution of outcomes is introduced
\begin{align}
	P_\textrm{eff}(\obs) = \frac{P_+(\obs)^{\optPar^*} P_-(\obs)^{1-\optPar^*}}{\int d\obs\, P_+(\obs)^{\optPar^*} P_-(\obs)^{1-\optPar^*}}.
\end{align}
Finally, the value of $K_-''(\optPar^*)$ is obtained as the second cumulant of the log-likelihood ratio with respect to the effective distribution:
\begin{align}
	K_-''(\optPar^*) = \int d\obs\, P_\textrm{eff}(\obs) \LLR(\obs)^2.
\end{align}
Note that higher derivatives $K_-^{(n)}(\optPar^*)$ are similarly obtained as the higher cumulants of $\LLR(\obs)$ with respect to $P_\textrm{eff}(\obs)$.

\subsection{Generalized central limit}

The Chernoff information is a distance measure between the two readout noise distributions $P_\pm(\obs)$. It follows that $P_+(\obs)\approx P_-(\obs)$ in the limit $\CI \rightarrow 0$. In that limit, it is convenient to rewrite the distributions as
\begin{align}
	P_\pm(\obs) = \bar{P}(\obs)\left[1 \pm \frac{y(\obs)}{2}\right].
\end{align}
Here, $y$ is the relative error between the distributions,
\begin{align}
	y(\obs) = \frac{\delta P(\obs)}{\bar{P}(\obs)},
\end{align}
and $\bar{P}(\obs)$ and $\delta P(\obs)$ are the average and difference distributions, respectively,
\begin{align}
\begin{split}
	&\bar{P}(\obs) = \frac{P_+(\obs) + P_-(\obs)}{2}, \\
	&\delta P(\obs) = P_+(\obs) - P_-(\obs).
\end{split}
\end{align}
The cumulant-generating function $K_-(\optPar)$, Eq.~\eqref{eq:CGFLLRExplicit}, may be expanded in powers of $y$. Performing the expansion and minimizing $K_-(\optPar)$ to leading order, it is found that the leading contributions to $K_-(\optPar)$ and its derivatives at the optimum are
\begin{align}
\begin{split}
	&K_-(\optPar^*) \approx -\left<y^2\right>/8, \\
	&K_-''(\optPar^*) \approx \left<y^2\right>, \\
	&K_-^{(3)}(\optPar^*) \approx \left<y^3\right>, \\
	&K_-^{(4)}(\optPar^*) \approx \left<y^4\right> - 3\left<y^2\right>^2,
\end{split}
\end{align}
where the expectation values are taken with respect to the average distribution $\bar{P}(\obs)$. Note that the higher derivatives are cumulants of the relative error $y(\obs)$ with respect to $\bar{P}(\obs)$. In addition, the parameters $\ngPar$ and $\optPar^*$ are approximately
\begin{align}
\begin{split}
	&\ngPar \approx 1 + \frac{1}{16}\left<y^2\right> + \frac{1}{48}\frac{\left<y^3\right>^2}{\left<y^2\right>^2} - \frac{1}{48}\frac{\left<y^4\right>}{\left<y^2\right>}, \\
	&\optPar^* \approx \frac{1}{2} + \frac{1}{24}\frac{\left<y^3\right>}{\left<y^2\right>}.
\end{split}
\end{align}
These expressions show that $\ngPar \rightarrow 1$ and $\optPar^* \rightarrow 1/2$ as $\CI$ (and thus $y$) approach zero. They may be rewritten as
\begin{align}
\begin{split}
	&\ngPar \approx 1 + \frac{1}{48}\left[\left(\frac{K_-^{(3)}(\optPar^*)}{K_-''(\optPar^*)}\right)^2 - \frac{K_-^{(4)}(\optPar^*)}{K_-''(\optPar^*)}\right], \\
	&\optPar^* \approx \frac{1}{2} + \frac{1}{24}\frac{K_-^{(3)}(\optPar^*)}{K_-''(\optPar^*)}.
\end{split}
\end{align}
Therefore, the deviations of $\ngPar$ from unity and of $\optPar^*$ from $1/2$ are controlled by the ratio of the higher cumulants to the second cumulant. This makes it clear that $\ngPar$ and $\optPar^*$ measure the deviation of the log-likelihood ratio $\cumLLR_N$ from Gaussian behavior.

\section{MONTE CARLO SIMULATIONS \label{app:simulations}}

\subsection{Hidden Markov model}

The results of Sec.~\ref{sec:generalized} are obtained by simulating the cumulative readout error rate $\cumErr_N$ for arbitrary distributions of analog readout outcomes and in the presence of non-QND imperfections. In the presence of non-QND imperfections, the eigenvalue $a$ may change from one repetition to the next. For Markovian non-QND imperfections in the single-shot readout regime, the statistics of these transitions are effectively classical (see Appendix~\ref{app:qndReadout}). Therefore, the dynamics are fully characterized by the probability $P_{a_N}(a_{N+1})$ to transition to eigenvalue $a_{N+1}$ given the previous eigenvalue $a_N$~\footnote{Here, the state space is two-dimensional, $a = \pm 1$. In general, however, the measured binary observable $A$ may be embedded in a larger state space with more complex dynamics.}. Moreover, the distribution of analog readout outcomes $P_{a_N}(\obs_N)$ is now conditioned on the eigenvalue $a_N$ realized in repetition $N$. In Sec.~\ref{sec:universalityNonQND}, the common case where the observable eigenvalue relaxes from $a_k = +1$ to $a_{k+1} = -1$ at rate $\Gamma$ is considered. For this relaxation process, the transition probabilities $P_{a_k}(a_{k+1})$ in a repetition of duration $\Delta t$ are
\begin{align}
\begin{array}{ll}
	P_{a_k = +1}(a_{k+1} = +1) \approx 1-p, & P_{a_k = -1}(a_{k+1} = +1) = 0 \\
	P_{a_k = +1}(a_{k+1} = -1) \approx p, & P_{a_k = -1}(a_{k+1} = -1) = 1,
\end{array}
\end{align}
where $p = \Gamma \Delta t \ll \min(\CI,1)$ is the transition probability. Processes described by the distributions $P_{a_N}(a_{N+1})$ and $P_{a_N}(\obs)$ define a hidden Markov model~\cite{zucchini2009}. Such models can be sampled and decoded efficiently as described below.

\subsection{Sampling}

A large number $M = 10^6$ of independent strings of readout outcomes $\Obs_N = \left\{\obs_0,\obs_1,\dots,\obs_{N-1}\right\}$ is sampled for both initial eigenvalues $a_0 = \pm 1$. Each string is sampled with the following algorithm:
\begin{enumerate}
	\item Set the initial eigenvalue $a_0$ and set $k=0$;
	\item Repeat the following until $k=N$:
	\begin{enumerate}
	\item Sample the readout outcome $\obs_k$ from the distribution $P_{a_k}(\obs_k)$;
	\item Sample the next eigenvalue $a_{k+1}$ according to the distribution $P_{a_k}(a_{k+1})$;
	\item Increase $k$ by $1$.
	\end{enumerate}
\end{enumerate}

\subsection{Decoding}

Decoding is performed by determining which initial eigenvalue $a_0$ most likely generated the sampled data. As discussed in Sec.~\ref{sec:multipleRepetitions}, this is done by calculating the log-likelihood ratio
\begin{align}
	\cumLLR_N = \ln \frac{P_{a_0=+1}(\Obs_N)}{P_{a_0=-1}(\Obs_N)},
\end{align}
where $P_{a_0}(\Obs_N)$ are the probabilities of obtaining the string $\Obs_N$ conditioned on the initial eigenvalue $a_0$. The probabilities $P_{a_0}(\Obs_N)$ are the likelihoods for the eigenvalues $a_0$. If $\cumLLR_N > 0$, $a_0 = +1$ is assigned. If $\cumLLR_N < 0$, $a_0 = -1$ is assigned. If $\cumLLR_N = 0$, the value of $a_0$ is assigned at random. If the assigned value of $a_0$ differs from the true value, an error has occurred. For each true value $a_0 = \pm 1$, the number of errors $\mathcal{E}_{\pm,N}$ is divided by the number of simulations $M$ to yield an estimate $\cumErr_{\pm,N} \approx \mathcal{E}_{\pm,N}/M$ of the cumulative error rate. The statistical uncertainty in that estimate is
\begin{align}
\delta \cumErr_{\pm,N} \approx \sqrt{\frac{\cumErr_{\pm,N}(1-\cumErr_{\pm,N})}{M}}.
\end{align}
Moreover, the statistical uncertainty in the average error rate $\cumErr_N = \left(\cumErr_{+,N} + \cumErr_{-,N}\right)/2$ is
\begin{align}
\delta \cumErr_N \approx \frac{1}{2}\sqrt{\delta \cumErr_{+,N}^2 + \delta \cumErr_{-,N}^2}.
\end{align}

\subsection{Calculation of the likelihood}

The likelihoods $P_{a_0}(\Obs_k)$ can be efficiently calculated for all substrings $\Obs_k$, $k \leq N$, using the procedure described below. In what follows, the dependence on $a_0$ is omitted to simplify notation. The likelihood can be decomposed as
\begin{align}
	P(\Obs_k) = \sum_{a_k} \ell_k(a_k),
\end{align}
where
\begin{align}
	\ell_{k}(a_k) \equiv P(\Obs_k,a_k).
\end{align}
The advantage of this decomposition is that $\ell_k(a_k)$ can be calculated iteratively using the theory of hidden Markov models~\cite{gambetta2007,myerson2008,gammelmark2014,ng2014,wolk2015,danjou2016,hann2018,nakajima2019,bultink2020,yoneda2020,xue2020-2,martinez2020}. Let $\bs{\ell}_k$ be a column vector with elements $\ell_k(a_k)$ in the basis $\left\{a_k = +1,a_k = -1\right\}$. This vector obeys the recurrence relation~\cite{zucchini2009}
\begin{align}
	\bs{\ell}_{k+1} = \bs{V}_k\cdot\bs{\ell}_k.
\end{align}
Here, $\bs{V}_k$ is a matrix with elements that depend on $\obs_k$:
\begin{align}
	V_k(a_{k+1},a_k) = P_{a_k}(\obs_k)P_{a_k}(a_{k+1}).
\end{align}
The probability of the string $\Obs_k$ occurring is then
\begin{align}
	P(\Obs_k) = \tr\left[\bs{\ell}_k\right].
\end{align}
Here, the trace of a vector is defined as the sum of its elements. Note that the above recurrence automatically yields the likelihood for all $k\leq N$ after $N$ iterations. The initial state is set to $\bs{\ell}_0 = (1,0)^T$ to calculate the likelihood for $a_0 = +1$ and to $\bs{\ell}_0 = (0,1)^T$ to calculate the likelihood for $a_0 = -1$. A numerically stable algorithm to calculate the log-likelihood is summarized below:
\begin{enumerate}
	\item Set $\bs{p}_0 = \bs{\ell}_0$ and set $k=0$;
	\item Repeat the following until $k=N$:
	\begin{enumerate}
	\item Calculate $\tilde{\bs{p}}_{k+1} = V_k \cdot \bs{p}_k$;
	\item Calculate $\mathcal{N}_{k+1} = \tr\left[\tilde{\bs{p}}_{k+1}\right]$;
	\item Update the normalized vector as $\bs{p}_{k+1} = \tilde{\bs{p}}_{k+1}/\mathcal{N}_{k+1}$;
	\item Update the log-likelihood as $\ln P(\Obs_{k+1}) = \ln P(\Obs_k) + \ln \mathcal{N}_{k+1}$;
	\item Increase $k$ by $1$.
	\end{enumerate}
\end{enumerate}
This update procedure is the direct classical analog of Eq.~\eqref{eq:quantumOperation}.

\section{CALCULATION OF THE SOFT-DECODING ADVANTAGE \label{app:softDecodingAdvantage}}

\subsection{Chernoff information for binary readout outcomes}

A hard-decoding strategy converts each analog outcome $\obs$ to binary outcomes $\pm$ in each repetition. The conditioned probabilities for the binary outcomes are
\begin{align}
\begin{array}{ll}
	P_+(+) = 1-\err_+, &P_-(+) = \err_- ,\\
	P_+(-) = \err_+, &P_-(-)=1-\err_-.
\end{array} \label{eq:binaryDisribution}
\end{align}
Here, the $\err_\pm$ are the conditioned single-repetition error rates defined in Eq.~\eqref{eq:conditionedErrors}. The Chernoff information $\CI_b$ for binary outcomes is then obtained by substituting Eq.~\eqref{eq:binaryDisribution} into Eq.~\eqref{eq:chernoffInformation}. The optimization over $\optPar$ can be performed exactly. The result is
\begin{align}
	\CI_b = - \ln\left[(1-\err_+)^{\optPar^*} \err_-^{1-\optPar^*} + \err_+^{\optPar^*}(1-\err_-)^{1-\optPar^*} \right], \label{eq:binaryChernoffInformation}
\end{align}
where
\begin{align}
\begin{split}
	&\optPar^* = \frac{\ln \left[\frac{\left(1-\err_-\right)}{\err_-} \frac{\ln \left(\frac{1-\err_-}{\err_+}\right)}{\ln \left(\frac{1-\err_+}{\err_-}\right)}\right]}{\ln\left[\frac{(1-\err_+)(1-\err_-)}{\err_+ \err_-}\right]}, \\
	&1 - \optPar^* = \frac{\ln \left[\frac{\left(1-\err_+\right)}{\err_+} \frac{\ln \left(\frac{1-\err_+}{\err_-}\right)}{\ln \left(\frac{1-\err_-}{\err_+}\right)}\right]}{\ln\left[\frac{(1-\err_+)(1-\err_-)}{\err_+ \err_-}\right]}. \label{eq:binaryChernoffOptimum}
\end{split}
\end{align}
Simple expressions for $\CI_b$ can be obtained in two limits of practical relevance. When $\err_\pm \rightarrow 0$, Eq.~\eqref{eq:binaryChernoffOptimum} takes the form
\begin{align}
\begin{split}
&\optPar^* \sim \frac{\ln(\err_-^{-1})}{\ln(\err_+^{-1})+\ln(\err_-^{-1})}, \\
&1 - \optPar^* \sim \frac{\ln(\err_+^{-1})}{\ln(\err_+^{-1})+\ln(\err_-^{-1})}.
\end{split}
\end{align}
Substituting these expressions back into Eq.~\eqref{eq:binaryChernoffInformation} gives Eq.~\eqref{eq:asymptoticBinaryChernoffInformation}. Another case of interest is the symmetric case, $\err_+ = \err_- = \err$. In this case, $\optPar^* = 1/2$ and Eq.~\eqref{eq:binaryChernoffInformation} reduces to
\begin{align}
	\CI_b = \ln \left[\frac{1}{\sqrt{4\err(1-\err)}}\right]. \label{eq:binaryChernoffSymmetric}
\end{align}

\subsection{Soft-decoding advantage for Gaussian noise}

The Chernoff information for Gaussian noise is obtained by substituting Eq.~\eqref{eq:distsGaussian} into Eq.~\eqref{eq:chernoffInformation}. By symmetry of the distributions, the infimum occurs at $\optPar^*=1/2$. Performing the Gaussian integral yields
\begin{align}
	\CI = \frac{r}{2}. \label{eq:gaussianChernoff}
\end{align}
Next suppose that the analog outcomes are binarized in each repetition. By symmetry, it is clear that $\err_+ = \err_- = \err$. The average single-repetition error rate for Gaussian noise is simply
\begin{align}
\begin{split}
\err_G &= \int_{-\infty}^0 d\obs\, P_+(\obs) \\
&= \int_0^{\infty} d\obs\, P_-(\obs) = \frac{1}{2}\erfc\left(\sqrt{\frac{r}{2}}\right).
\end{split}
\end{align}
The Chernoff information for binary outcomes, Eq.~\eqref{eq:binaryChernoffSymmetric}, is then
\begin{align}
	\CI_b = \ln \left[\frac{1}{\sqrt{4\err_G(1-\err_G)}}\right]. \label{eq:binaryGaussianChernoff}
\end{align}
Expanding this expression in the two extreme limits $r \ll 1$ and $r \gg 1$ gives
\begin{align}
\CI_b =
\left\{
\begin{array}{lcl}
	\frac{r}{\pi} - \left(\frac{\pi-3}{3\pi^2}\right)r^2 + O\left(r^3\right) & \textrm{for} & r \ll 1,\\
	\frac{r}{4} + \frac{1}{4}\ln\left(\frac{\pi r}{8}\right) + O\left(\frac{1}{r}\right)& \textrm{for} & r \gg 1.
\end{array}
\right. \label{eq:binaryChernoffExpansion}
\end{align}
Comparing Eqs.~\eqref{eq:gaussianChernoff} and \eqref{eq:binaryChernoffExpansion} gives the soft-decoding advantage $\SDA = \CI/\CI_b$ in both limits:
\begin{align}
\SDA =
\left\{
\begin{array}{lcl}
	\frac{\pi}{2} + \left(\frac{\pi - 3}{6}\right)r + O\left(r^2\right) & \textrm{for} & r \ll 1,\\
	2 - \frac{2}{r}\ln\left(\frac{\pi r}{8}\right) + O\left[\left(\frac{1}{r}\ln r\right)^2\right] & \textrm{for} & r \gg 1.
\end{array}
\right.
\end{align}
The expression for $r \gg 1$ is the one obtained in Ref.~\cite{danjou2014-2} by other means.

\subsection{Soft-decoding advantage for Gaussian noise with conversion errors}

The Chernoff information for Gaussian noise with conversion errors is obtained by substituting Eq.~\eqref{eq:distsConversion} into Eq.~\eqref{eq:chernoffInformation}. By symmetry of the distributions, the infimum occurs at $\optPar^* = 1/2$. Rearranging the integral gives
\begin{align}
\begin{split}
	\CI = \frac{r}{2} - \ln\left[\int_{-\infty}^{\infty} dx\,\frac{e^{- \frac{x^2}{2}}}{\sqrt{2\pi}} \sqrt{1 + 4\eta(1-\eta)\sinh^2(\sqrt{r}x)} \right].
\end{split}
\label{eq:conversionChernoff}
\end{align}
The second term gives a correction to Eq.~\eqref{eq:gaussianChernoff} due to the finite rate of conversion errors $\eta$. The symmetry of the distributions also means that $\err_+ = \err_- = \err$. The average single-repetition error rate in the presence of conversion errors is
\begin{align}
	\err_\eta = (1-\eta) \err_G + \eta (1-\err_G),
\end{align}
where $\err_G = \erfc\left(\sqrt{r/2}\right)/2$ as before. The Chernoff information for binary outcomes is then obtained from Eq.~\eqref{eq:binaryChernoffSymmetric}:
\begin{align}
	\CI_b = \ln \left[\frac{1}{\sqrt{4\err_\eta(1-\err_\eta)}}\right]. \label{eq:conversionBinaryChernoff}
\end{align}
The soft-decoding advantage $\SDA = \CI/\CI_b$ can be calculated from Eqs.~\eqref{eq:conversionChernoff} and \eqref{eq:conversionBinaryChernoff} by performing a simple integral. The result is plotted as a function of $\err_G$ and $\eta$ in Fig.~\ref{fig:fig3}.

\bibliography{bibliography}

\end{document}